\documentclass{elsart}
\usepackage{graphicx}
\usepackage{amssymb}
\begin{document}
%
%
\begin{frontmatter}
\title{The $A(K^-_{stop},\pi^\pm\Sigma^\mp)A'$ reaction on p-shell nuclei} 
\vspace{-5mm}
\centering{\large FINUDA Collaboration}
\author[polito]{M.~Agnello}, 
\author[lnf]{L.~Benussi},
\author[lnf]{M.~Bertani},
\author[korea]{H.C.~Bhang},
\author[unibs]{G.~Bonomi},
\author[unitos]{E.~Botta}, 
\author[units]{M.~Bregant},
\author[unitos]{T.~Bressani},
\author[infnto]{S.~Bufalino},
\author[unitog]{L.~Busso},
\author[infnto]{D.~Calvo},
\author[units]{P.~Camerini},
\author[uniba]{B.~Dalena},
\author[unitos]{F.~De~Mori},
\author[uniba]{G.~D'Erasmo},
\author[lnf]{F.L.~Fabbri}, 
\author[infnto]{A.~Feliciello}, 
\author[infnto]{A.~Filippi},  
\author[uniba]{E.~M.~Fiore}, 
\author[infnpv]{A.~Fontana},
\author[kek]{H.~Fujioka},
\author[infnpv]{P.~Genova},
\author[lnf]{P.~Gianotti},  
\author[infnts]{N.~Grion}, 
\author[lnf]{V.~Lucherini},
\author[unitos]{S.~Marcello}, 
\author[teheran]{N.~Mirfakhrai},
\author[unibs] {F.~Moia}, 
\author[infnpv]{P.~Montagna},
\author[cnr]{O.~Morra}, 
\author[kek]{T.~Nagae}, 
\author[riken]{H.~Outa},  
\author[infnba]{A.~Pantaleo\thanksref{deceased}},  
\author[infnba]{V.~Paticchio}, 
\author[infnts]{S.~Piano\thanksref{corresponding}},   
\author[units]{R.~Rui}, 
\author[uniba]{G.~Simonetti}, 
\author[infnto]{R.~Wheadon}, 
\author[unibs]{A.~Zenoni}
\thanks[corresponding]{corresponding author. E-mail: stefano.piano@ts.infn.it; 
Fax:+39.040.5583350.}
\thanks[deceased]{deceased.} 
\address[polito]{Dip. di Fisica Politecnico di Torino, via Duca degli Abruzzi
Torino, Italy, and INFN Sez. di Torino, via P. Giuria 1 Torino, Italy}
\address[lnf]{Laboratori Nazionali di Frascati dell'INFN, via E. Fermi 40 
Frascati, Italy}
\address[korea]{Dep. of Physics, 
Seoul National Univ., 151-742 Seoul, South Korea}
\address[unitos]{Dipartimento di Fisica Sperimentale, Universit\`a di
Torino, via P. Giuria 1 Torino, Italy, and INFN Sez. di Torino, 
via P. Giuria 1 Torino, Italy} 
\address[units]{Dip. di Fisica Univ. di Trieste, via Valerio 2 Trieste, 
Italy and INFN, Sez. di Trieste, via Valerio 2 Trieste, Italy}
\address[unitog]{Dipartimento di Fisica Generale, Universit\`a di
Torino, via P. Giuria 1 Torino, Italy, and INFN Sez. di Torino, 
via P. Giuria 1 Torino, Italy} 
\address[infnto]{INFN Sez. di Torino, via P.  Giuria 1 Torino, Italy}
\address[uniba]{Dip. di Fisica Univ. di Bari, via Amendola 179 Bari, 
Italy and INFN Sez. di Bari, via Amendola 179 Bari, Italy }
\address[infnpv]{INFN Sez. di Pavia, via Bassi 6 Pavia, Italy}
\address[infnts]{INFN, Sez. di Trieste, via Valerio 2 Trieste, Italy}
\address[teheran]{Dep of Physics Shahid Behesty Univ., 19834 Teheran, Iran}
\address[cnr]{INAF-IFSI Sez. di Torino, C.so Fiume, Torino, Italy
and INFN Sez. di Torino, via P. Giuria 1 Torino, Italy} 
\address[kek]{Dep. of Physics, Kyoto University, Kyoto, 606-8502, Japan.} 
\address[riken]{RIKEN, Wako, Saitama 351-0198, Japan}
\address[infnba]{INFN Sez. di Bari, via Amendola 179 Bari, Italy }
\address[unibs]{Dip. di Meccanica, Universit\`a di Brescia, via Valotti 9 
Brescia, Italy and INFN Sez. di Pavia, via Bassi 6 Pavia, Italy}
\end{frontmatter} 
%
%
\setlength{\baselineskip}{2.3ex}          
{\small\bf Abstract:}
{
This letter is concerned with the study of the $K^-_{stop}A\rightarrow
\pi^\pm\Sigma^\mp A'$ reaction in p-shell nuclei, i.e., $^{6,7}Li$, $^9Be$, 
$^{13}C$ and $^{16}O$. The $\pi^\pm\Sigma^\mp / K^-_{stop}$ emission rates 
are reported as a function of $A$. These rates are discussed in comparison 
with previous findings. The ratio $\pi^-\Sigma^+/\pi^+\Sigma^-$ in p-shell 
nuclei is found to depart largely from that on hydrogen, which provides 
support for large in-medium effects possibly generated by the sub-threshold 
$\Lambda(1405)$. The continuum momentum spectra of prompt pions and free 
sigmas are also discussed as well as the $\pi^\pm\Sigma^\mp$ missing mass 
behavior and the link with the reaction mechanism. The apparatus used for 
the investigation is the FINUDA spectrometer operating at the DA$\Phi$NE 
$\phi$-factory (LNF-INFN, Italy).
}

\normalsize
%
%
\begin{center}
{\bf I. INTRODUCTION}
\end{center}
This letter describes an experimental study of the reaction $K^-_{stop}
~A\rightarrow\pi^\pm\Sigma^\mp A'$, where $A$ is $^6Li$, $^7Li$, $^9Be$, 
$^{13}C$ and $^{16}O$. In this study, the charged $\Sigma$-hyperons are 
reconstructed through their decay channels $\Sigma^\pm\rightarrow n\pi^
\pm$. The requirement of having one neutron and two charged pions in the 
final state excludes charged $\Sigma$'s and neutral $\Lambda$'s from the 
two-body absorption process $K^-_{stop}~2N\rightarrow N\Sigma(\Lambda)$ 
since both final channels are characterized by a single pion. Moreover, 
the request for a final $\Sigma$ excludes detection of the $\Sigma$ 
conversion reaction, $\Sigma N\rightarrow\Lambda N$, and thus the 
identification of possible $\Sigma$-hypernuclear states. For these 
reasons, this measurement studies the elementary process $K^-_{stop}~p
\rightarrow\pi^\pm\Sigma^\mp$ with the absorbing proton being embedded 
in $A$.

To understand the $K^-$ nuclear and the $Y$ nucleon(s) interactions, 
FINUDA pursued a program of $K^-_{stop}~A$ experimental studies through 
the methodical analysis of the reaction channels. This was to determine 
the $A$-dependence of $\Lambda$ capture into hypernuclear bound states 
\cite{expt:agnello1,expt:agnello2}, which helped to constrain the 
threshold  $K^-$ nuclear potential \cite{theor:cieply}. Study of the 
mesonic \cite{expt:agnello3} and non-mesonic \cite
{expt:agnello4,expt:agnello5} 
weak decay of p-shell $\Lambda$-hypernuclei was also part of the $K^-
_{stop}~A$ studies since it helps in understanding the $\Lambda N$ weak 
interaction. The analyzes of the pion-less $K^-$ absorption leading 
to the selected $\Lambda p$ \cite{expt:agnello6}, $\Lambda d$ \cite
{expt:agnello7} and $\Lambda t$ \cite{expt:agnello8} channels established 
the possibility of $K^-$'s forming quasi-bound states. This is still a 
debated topic; however, it has spurred widespread interest \cite
{theor:weise-gal}. The presence of nuclear matter may sensibly  
modify  the sub-threshold behavior of the $\overline{K}$-N interaction 
depending on the nuclear density 
\cite{theor:wycech,theor:staronski,theor:ohnishi,theor:friedman}. This 
occurrence may favorably be probed with the $K^-_{stop}A\rightarrow\pi
\Sigma A'$ reaction since the $\pi\Sigma$ channel prevails over the other 
competing channels. Furthermore, the measurement involves a single proton 
bound in $A$ and takes place at threshold, which also facilitates the 
study of the under-threshold behavior of the $\overline{K}$-N interaction.

Kaons, after stopping in nuclei, initially form K-atoms, leading to kaon 
absorption by nucleons lying at the nuclear surface, giving rise to a 
number of quasi-free reactions. This analysis selectively examines the 
$K^-_{stop}~p\rightarrow\pi^\pm\Sigma^\mp$ processes, where $p$ is a 
bound proton. The prompt pion and free sigma momentum distributions 
describe the $\pi$- and $\Sigma$-continuum behavior in the reaction $K^-_
{stop}~A\rightarrow\pi^\pm\Sigma^\mp A'$. These spectra are discussed in 
detail rather than the $\pi^\pm\Sigma^\mp$ invariant masses. In this case, 
the $\pi\Sigma$ channel is filled by two resonances $\Sigma(1385)$ and 
$\Lambda(1405)$ as well as the $\pi\Sigma$ continuum whose phase space 
develops in the same region as the two resonances \cite{theor:oset}. The 
effects of $\pi$-$A'$ and $\Sigma$-$A'$ final-state interactions further 
complicate this scenario since they distort the initial shape of invariant 
mass distributions. The $\pi(\Sigma)$ momentum spectra may give a clearer 
picture of the dynamics of the $K^-_{stop}$ absorption by a bound 
$p$. The $\pi\Sigma$ missing mass spectra are discussed with the aim of 
measuring the energy lost via undetected particles. This is to investigate 
the number of nucleons involved in the kaon absorption and hence the 
reaction mechanism. The article also focuses on the $\pi^\pm\Sigma^\mp/
K^-_{stop}$ emission rates, which are compared to earlier bubble chamber 
measurements. Moreover, they are used to form the $R_{+-}$ ratio (i.e., 
$\pi^-\Sigma^+$/$\pi^+\Sigma^-$) which is known to change significantly
when moving the kaon absorption from a free proton to a bound proton
\cite{theor:wycech,theor:staronski,theor:ohnishi,theor:friedman}.
\begin{center}
{\bf II. THE EXPERIMENTAL PROCEDURE}
\end{center}
All the particles involved in the $A(K^{-}_{stop},\pi^\pm\Sigma^\mp)A'$ 
reaction, i.e., $K^{-}$'s (and correlated $K^+$'s), $\pi^\pm$'s and 
$\Sigma^\mp$'s (from $n \pi^\mp$ decays), are fully reconstructed.
Negative and positive kaons result from  decays of phi meson, $\phi
\rightarrow K^-K^+$ (B.R.$\sim$ 50\%), $\phi$'s produced by the $e^+e^-
\rightarrow\phi(1020)$ reaction at the DA$\Phi$NE collider. The $\phi$ 
mesons are created nearly at rest giving rise to kaon pairs which are 
emitted in nearly opposite directions with a kinetic energy of 16.1$\pm
$1.5 MeV. The kaon pairs, after traversing the 2 inner layers of the 
spectrometer, stop in $\sim$0.25 gr/cm$^2$ thick targets about 20.0$
\times$5.3 cm$^2$ in area. Most are solid slabs except for $^{13}C$ and 
$^{16}O$, the first of which is carbon powder and second liquid $D_2O$. 
When a $K^-$ starts the reaction, the $K^-K^+$ pair triggers the inner 
layer of FINUDA and, simultaneously, a final state particle must be 
detected by the outer layer of FINUDA. Such a selective trigger is 
mandatory to reject the overwhelming background originated by electrons 
and positrons circulating in DA$\Phi$NE.

The inner (TOFINO \cite{FINUDA:00}) and outer (TOFONE \cite{FINUDA:01})
sensitive layers of FINUDA are two segmented detectors made of plastic 
scintillator, which are  organized as the staves of a barrel. The TOFINO 
barrel, which encloses the $e^+e^-$ interaction volume, has a diameter of 
11 cm and is used for starting the time-of-flight system ($tof$) and for 
triggering purposes. TOFONE, which consists of 72 trapezoidal slabs 255 
cm long and 10 cm thick, is the stop-counter of the FINUDA $tof$, which 
is used to determine the momentum of neutrons as well as for triggering 
purposes. The neutron momentum resolution of FINUDA is obtained from the 
distribution width of monokinetic neutrons arising from the decays of 
stopped positive sigmas. The momentum distribution of such neutrons is 
shown in Fig. 1(a), which displays a peak at 187.6$\pm$0.2 MeV/c and a 
width of 8.4$\pm$0.2 MeV/c at $\sigma$. This width is slightly narrower 
than the one reported in a previous analysis 9.4$\pm$0.2 MeV/c \cite
{expt:agnello9} due to a finer calibration of the present FINUDA $tof$. 
Neutrons from the first run of FINUDA cannot be utilized for this analysis 
because of the high threshold chosen for TOFONE, which was mainly used for 
triggering purposes. As a result, neutrons below 20 MeV were hardly 
%
%
\begin{figure}[t,c,b]
 \centering
  \includegraphics*[width=1.0\textwidth]
    {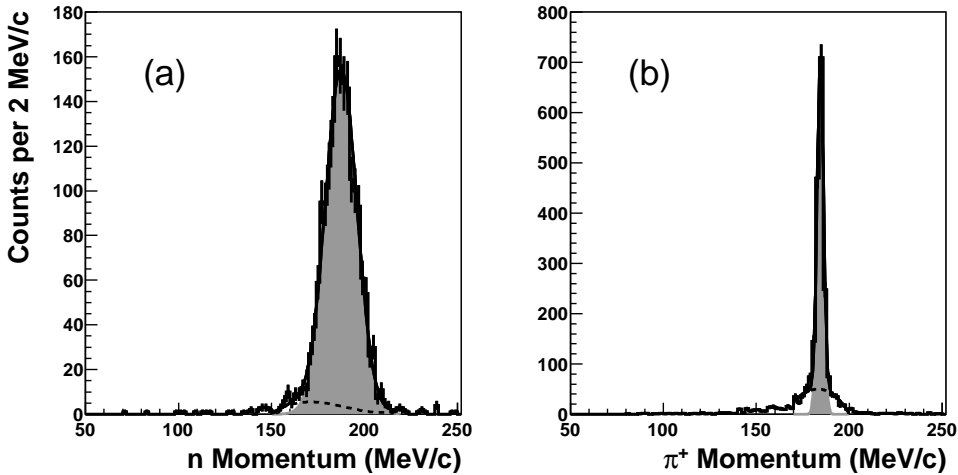}
    \caption{\footnotesize  Measured momentum distributions of neutrons 
      and positive pions from positive sigma decays at rest, 
      $\Sigma^+\rightarrow n\pi^+$. The grey-filled peaks are the result 
      of a Gaussian peak fitting, while the dotted lines are a polynomial 
      best-fit to the background.}
\end{figure}      
detected and hence sigmas could not actually be reconstructed. The tracking 
system of FINUDA is located between TOFINO and TOFONE. It consists of a 
vertex detector \cite{FINUDA:02}, two intermediate layers of low-mass drift 
chambers \cite{FINUDA:03} and an outer array of straw tubes \cite{FINUDA:04}. 
The vertex detector comprises 2 layers of double-sided micro-strip silicon 
sensors \cite{FINUDA:02} whose internal layer is located at about 7 cm from 
the trajectories of the $e^+e^-$ crossing beams. The 2 layers closely 
surround a stack of 8 solid targets. The outer six layers of straw tubes 
\cite{FINUDA:04} has the inner layer set at 111.0 cm from the crossing beams. 
With the nominal field set at 1 T, the spectrometer is capable of analyzing 
184.5$\pm$0.1 MeV/c  positive pions with a resolution at $\sigma$ of 1.72$
\pm$0.06 MeV/c (see Fig. 1(b)), which only slightly worsens for pions of 
lower momenta. However, the reduced spectrometer resolution only weakly 
alters the continuous behavior of pion (and sigma) spectra. 

%
%
\begin{figure}[t]
 \centering
  \includegraphics*[width=1.0\textwidth]
    {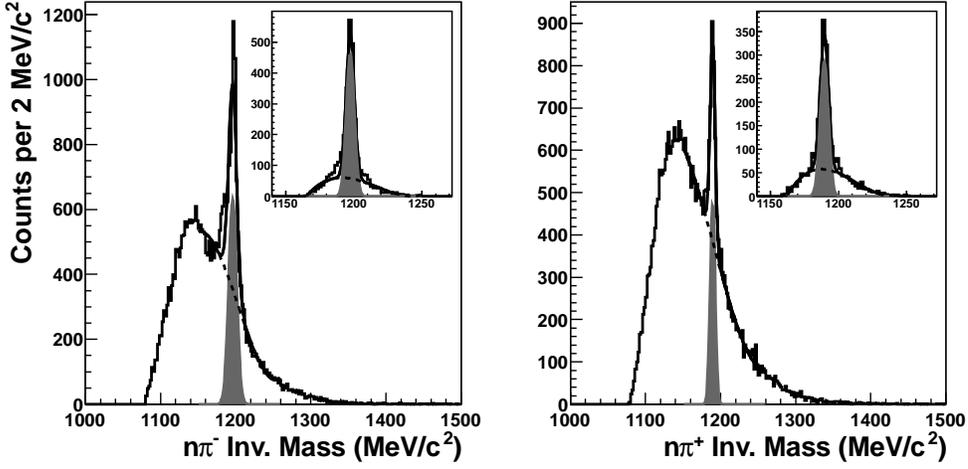}
    \caption{\footnotesize Unconstrained $n\pi^-$ and $n\pi^+$ invariant 
        mass distributions for the $^6Li(K^-_{stop},n\pi^\mp)A'$ reaction.  
	The grey-filled peaks represent the $\Sigma^-$ (left) and $\Sigma^+$ 
	signals. In the insets the constrained  $n\pi^\mp$ invariant 
	mass distributions are shown. The applied constraints are detailed 
	in the text.}
\end{figure}      
In the present analysis, both kinematic and topological constraints are 
applied to measured observables. The primary condition for processing a 
$n\pi^-\pi^+$ event requires a vertex in one of the targets between the 
trajectory described by the initial kaon pair and the track of one of 
the two final pions. A vertex position is reconstructed with an average 
spatial resolution of 0.7 mm, which is mostly due to the kaon straggling 
inside the target. In fact, the average resolution of the vertex detector in 
reconstructing a particle prong is 0.2 mm. The mass of a particle is then
defined by the particle energy deposit ($dE/dX$ technique) in the two 
layers of the vertex detector as well as in the two drift chambers. The 
$dE/dX$ technique ensures an overall pion ID efficiency above 98\%. Fig. 2 
shows the $n\pi^-$ and $n\pi^+$ invariant mass behaviors for $^6Li$. The 
sharp peaks identify the $\Sigma^\mp$ signals, which are placed at 1197.0$
\pm$3.4($\sigma$) MeV/c$^2$ and 1189.2$\pm$3.5($\sigma$) MeV/c$^2$, 
respectively. The signal to background ratio assigned by a peak-fitting 
procedure is $(S/B)_{\Sigma^{-}}$=1.0 and $(S/B)_{\Sigma^{+}}$=0.7, where 
$S$ is the area of the grey-filled peaks taken at 2$\sigma$ and $B$ is the 
area below the dotted curves taken in the same mass range. For these spectra, 
the background $B$ is mostly due to gammas which may emulate neutrons in 
TOFONE, and by neutrons which may scatter before being detected. The 
background due to misidentification of pions is negligible. A clear 
background reduction is obtained by requiring the kinematic observables of 
the $n\pi^\mp$ events to belong to the phase space volume of the $\Sigma^
\mp\rightarrow n\pi^\mp$ decays, which safely includes the prompt pions. As 
an example, when applying the above phase space constraints to $n\pi^\mp$ 
events, the $n\pi^-$ and $n\pi^+$ invariant mass distributions are dominated 
by $\Sigma^-$ and $\Sigma^+$ peaks, see the insets of Fig. 2. When peak 
fitting these distributions $(S/B)_{\Sigma^-}$=4.5  and  $(S/B)_{\Sigma^+
}$=3.6 thereby permitting a reliable discussion of the $\Sigma$ physics.  
\begin{center}
\bf {III. THE RESULTS}
\end{center}   
Prompt pions and free sigmas (see Fig. 2 and related discussion) are fully
reconstructed, which ensures a quasi-exclusive character to this measurement. 
Accordingly, the observables involved in the $\pi^\pm\Sigma^\mp$ analysis, 
i.e., missing masses, momentum distributions and emission rates, should  be
scarcely affected by background. Fig. 3 shows the ($K^-_{stop},\pi^\pm\Sigma
^\mp$) missing mass ({\it M}) spectra of $^6Li$. These spectra are formed 
%
%
\begin{figure}[b,c,t]
 \centering
  \includegraphics*[width=1.0\textwidth]
    {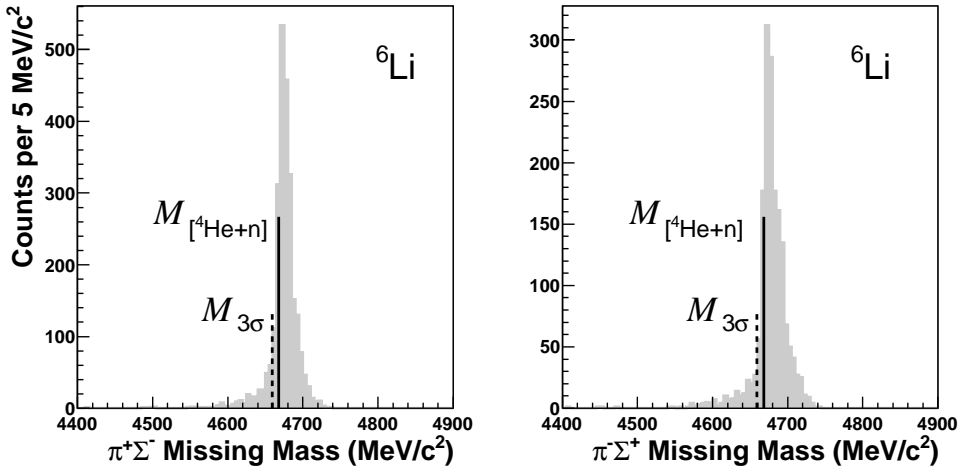}
    \caption{\footnotesize Missing mass distribution for the 
     $^6Li(K^-_{stop},\pi^\pm\Sigma^\mp)A'$ reactions. The full
     line denotes the physical threshold of the missing mass, the 
     dotted line indicates the lower threshold chosen for the 
     analysis.}
\end{figure}   
with the same data set of the constrained  $n\pi^\pm$ invariant masses, i.e.,
the grey-filled distributions in the insets of Fig. 2. The full line denotes 
the physical threshold of the missing mass, that is, the value acquired by 
{\it M} when $A'\equiv$ $[^4He_{g.s.}+n]$. The dotted line points to the 
missing mass threshold ({\it M}$_{3\sigma}$) used in this analysis, where 
the overall uncertainty in measuring {\it M} is $\sigma$=3.1 MeV/$c^2$, 
or a full-width at half-maximum FWHM=7.3 MeV/$c^2$. Both distributions are 
peaked at about 10 MeV/c$^2$ above {\it M}$_{[^4He+n]}$ and have a 
FWHM in the range 22-24  MeV/c$^2$.

%
%
\begin{figure}[b]
  \hspace{-10mm}
   \includegraphics*[width=1.08\textwidth]
    {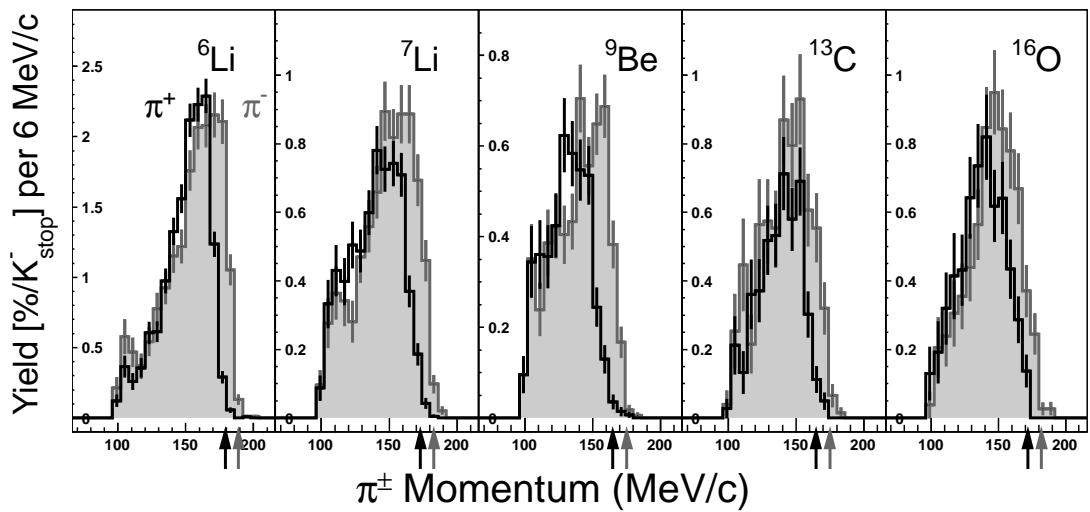}
    \caption{\footnotesize Measured momentum distributions of prompt 
      pions from the $A(K^-_{stop},\pi^\pm\Sigma^\mp)A'$ reactions. The 
      open (grey-filled) histograms describe the behavior of positive 
      (negative) pions and the black (grey) arrows point to the 
      p$[B_{\Sigma^{-(+)}}=0]$ threshold.}
\end{figure}      
Fig. 4 shows the measured momentum distributions of prompt pions from the 
$K^-_{stop}A\rightarrow\pi^\pm\Sigma^\mp A'$ reactions. The arrows denote 
the condition p$[B_{\Sigma^{-(+)}}=0]$, where $B_{\Sigma}$ is the binding
energy of possible $\Sigma$-hypernuclei ($_{\Sigma}A$). Therefore, broad 
peaks arising at p$[B_{\Sigma^{-(+)}}>0]$ would disclose the presence of 
$_{\Sigma}A$ states. Such states cannot be observed in the present spectra
since they are inhibited by requiring a $\Sigma$ to be present in the 
reaction exit channel. The distributions of pion (sigma in Fig. 5) momenta 
are corrected for the spectrometer acceptance, which is the largest source 
of systematic uncertainty. Such corrections relied on a Monte-Carlo 
code which embodied the FINUDA geometry. The code was required to generate 
$K^-_{stop}~A\rightarrow\pi^+\pi^- n~A'$ events with the final particles 
distributed uniformly over a range of momenta required by the reaction 
kinematics. The generated events fully crossing the geometry of FINUDA 
were finally reconstructed with the analysis code. The ratio of the 
reconstructed to generated events defines the acceptance of the apparatus. 
For a selected momentum bin, the systematic uncertainty of the acceptance 
results from the statistical uncertainty of the reconstructed events and 
the systematic uncertainty arising from the reconstruction of real events,
the two uncertainties being added in quadrature. Finally, the error bars 
associated with the data-points in Figs. 4 and 5 account for the statistics 
of real events and systematic uncertainty of the acceptance, again 
calculated by summing in quadrature the two uncertainties. The pion 
threshold of the spectrometer is slightly below 80 MeV/c; however, the pion
momentum  distributions are cut at 100 MeV/c to avoid large bin fluctuations 
due to acceptance correction. Note that above the p$[B_{\Sigma^{-(+)}}=0]$ 
thresholds the spectra are scarcely affected by background.  
%
%
\begin{figure}[t,c,b]
 \centering
  \includegraphics*[width=1.05\textwidth]
   {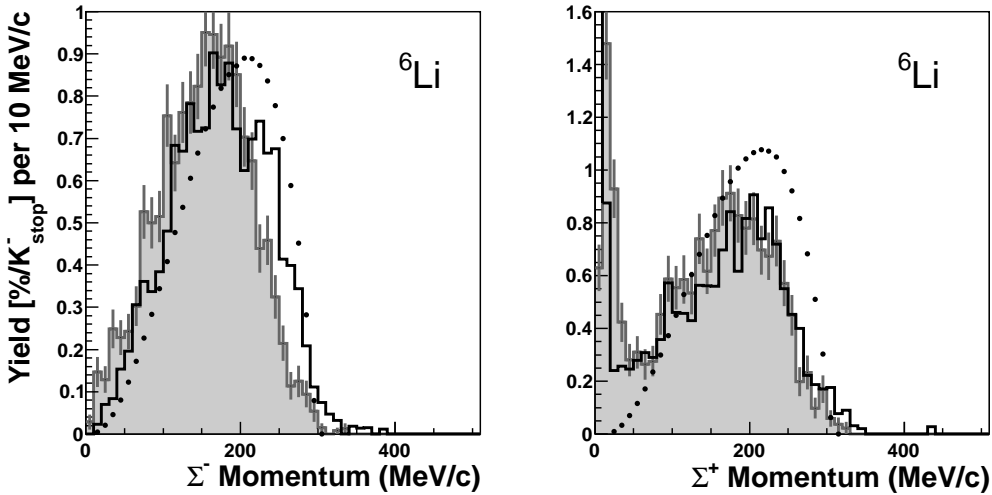}
    \caption{\footnotesize  Momentum distributions of sigmas from the 
      $^6Li(K^-_{stop},\pi^\pm\Sigma^\mp)A'$ reactions. The grey-filled 
      histograms are the measured distributions. The distributions of 
      Monte-Carlo generated sigmas are depicted by full dots, and with 
      open diagrams are represented the M-C generated sigmas being 
      reconstructed by FINUDA.}
\end{figure}      
The momenta of $\Sigma^\pm$'s are determined by measuring the momenta 
of their decay products. For $^6Li$, the measured momentum distributions 
of $\Sigma^\pm$-hyperons are shown in Fig. 5 (grey-filled histograms). For 
$\Sigma^-$, the distribution resembles a rather symmetric bump peaked at 
about 165 MeV/c with a FWHM of 155 MeV/c and shows no strength at around 0 
MeV/c. The $\Sigma^+$ momentum distribution  is instead peaked at around 0 
MeV/c and a 160 MeV/c broad-bump arises at $\sim$ 174 MeV/c. In this case, 
the FWHM of the bump is assessed by removing via a global-fitting method 
the 0 MeV/c peak, which ultimately smears out the line-shape of the bump. 
The $\Sigma^\pm$ bumps display little variation in shape: the $\sim$10 
MeV/c peak shift is explained by the different sigma masses, while the 
$\sim$5 MeV/c wider bump of the $\Sigma^+$-hyperons is well within the 
uncertainties in assessing the $\Sigma^\pm$ bump widths. For the other 
nuclei, the $\Sigma^\pm$ momentum distributions are not shown since they 
display shapes similar to the $\Sigma^\pm$ momentum distributions of $^
6Li$ with the only exception being the number of $\Sigma^+$ events in the 
0 MeV/c peak, which basically depends on the target size and density. Such 
a behavior suggests a way for representing all the $\Sigma^\pm$ measured 
momentum distributions, which is based on the fact that their symmetric 
line-shapes can be fitted by means of Gaussian functions. With this 
approach, the $\Sigma^+$ momentum distributions are found to have a mean 
value of 167$\pm$7 MeV/c and a FWHM 178$\pm$20 MeV/c, and the $\Sigma^-$ 
momentum distributions a mean value of 157$\pm$11 MeV/c and a FWHM 
161$\pm$9 MeV/c.

The sigma momentum distributions  appear rather dissimilar at around 0 
MeV/c. At these momenta, positive and negative sigmas behave differently 
when traveling through nuclear targets of finite thickness. For example, 
a $\Sigma^+$ can stop in a target without being absorbed by a surrounding 
nucleus, which explains the appearance of the peak at around 0 MeV/c. 
When a $\Sigma^-$ comes at rest, it is captured by a target nucleus and 
later undergoes a $\Sigma$-$\Lambda$ conversion. $\Lambda$-hyperons are 
not presently reconstructed hence the momentum distribution shows no 
strength at around 0 MeV/c. To account for the different behavior of 
low-energy charged sigmas in targets of finite size, a Monte-Carlo code 
was developed which employs the $K^-_{stop}A\rightarrow\pi^\pm\Sigma^\mp 
A'$ quasi-free reaction to generate the sigma momentum spectra. Then, the 
spectra were filtered through the geometry of FINUDA, which also accounts 
for the interaction of $\Sigma^\mp$ with the target media. The resulting 
momentum spectra (open histograms) are finally compared with the measured 
momentum spectra and show a reasonable overall agreement. The M-C generated 
$\Sigma^\mp$ momentum spectra are displayed in Fig. 5 with full dots. 

The $K^-_{stop}~A$ emission rate leading to final $\pi\Sigma$ pairs is 
defined by the ratio:
\begin{equation}
 R_{A} = \frac{N_{\pi\Sigma}}{N_{K_{stop}}}                 
\end{equation}
where $N_{\pi\Sigma}$ is the number of $\pi\Sigma$ pairs produced when 
$N_{K_{stop}}$ negative kaons stop in $A$. These two observables can be 
related to the number of $\pi\Sigma$ ($N^{CE}_{\pi\Sigma}$) and $K^-_
{stop}$ ($N^{CE}_{K_{stop}}$) counted events ($CE$) through the equations:
\begin{equation}
N_{\pi\Sigma}=\frac{N^{CE}_{\pi\Sigma}}{\epsilon_{\pi\Sigma}} 
\hspace{+0.5cm}   
and 
\hspace{+0.5cm}
N_{K_{stop}} = \frac{N^{CE}_{K_{stop}}}{\epsilon_{K_{stop}}}
\end{equation}
where the parameters $\epsilon$ correct the observables $N^{CE}_{\pi
\Sigma}$ and $N^{CE}_{K_{stop}}$ for instrumental flaws, namely, the 
apparatus finite acceptance, inability to fully reconstruct all the $n\pi
^\pm$ events, misidentification of particles and trigger inefficiencies. 
Both $\epsilon_{\pi\Sigma}$ and $\epsilon_{K_{stop}}$ are assessed with 
standard Monte-Carlo techniques. A second partially-independent approach 
to calculate both $N_{\pi\Sigma}$ and $N_{K_{stop}}$ is also available, 
which requires a muon from $K^+\rightarrow\mu^+\nu_\mu$ (BR=63.6\%) in 
coincidence with the $n\pi^+\pi^-$ initial event. Such an approach avoids 
any trigger inefficiency but also reduces the useful data-set by at least 
BR, which causes the statistical uncertainties to augment. When taken with 
their overall uncertainties, these emission rates are consistent with those 
calculated by using eqn (1), so only $R_{A}$'s are reported in Table 1. 

%
\begin{table}[b]
\caption[Table]
{The table lists $R_{A}$ for $\pi^-\Sigma^+$ and $\pi^+\Sigma^-$ as 
a function of $A$, where $R_{A}$ is given in units of [$10^{-2} / 
K^-_{stop}$]. The $R_{A}$ values are followed by the statistical and 
systematic uncertainties. $\Sigma^-_{loss}$ is expressed in units of 
[\%]. The emission rates of $^4He$ and $^{12}C$ are taken from Refs. 
\cite{expt:katz} and \cite{expt:vander-velde}, respectively. 
$R_{+-}$=$R_{A}(\pi^-\Sigma^+) / R_{A}(\pi^+\Sigma^-)$.}
\vspace{2.0mm}
\begin{center}    
\begin{tabular}{lc|c|c|c|c}                                               \hline\hline  
   \multicolumn{1}{c}  {$A$}                          &   &   
   \multicolumn{2}{c|} {$R_{A}$}                          &
   \multicolumn{1}{c|} {$\Sigma^-_{loss}$}                &           
   \multicolumn{1}{c}  {$R_{+-}$}                                     \\      
& &\multicolumn{1}{c|} {$\pi^-\Sigma^+$}                  & 
   \multicolumn{1}{c|} {$\pi^+\Sigma^-$}                  &            \\ 
\hline
$^4He$  & &$12.8$$\pm 2.4$     &$ 7.2\pm 1.4$       &$-$&       1.8$\pm 0.5$       \\
$^6Li$  & &$17.6\pm 1.0\pm 1.3$&$14.4\pm 1.1\pm 0.2$&30$\pm 2$& 1.2$\pm$0.1$\pm$0.1\\
$^7Li$  & &$ 7.8\pm 0.5\pm 0.2$&$ 6.4\pm 0.4\pm 0.6$&26$\pm 2$& 1.2$\pm$0.1$\pm$0.1\\
$^9Be$  & &$ 5.5\pm 0.3\pm 0.3$&$ 4.3\pm 0.2\pm 0.5$&38$\pm 3$& 1.3$\pm$0.1$\pm$0.2\\
$^{12}C$& &$ 9.2\pm 0.3       $&$ 7.3\pm 0.2       $&$-$      & 1.25$\pm 0.09$     \\
$^{13}C$& &$ 6.8\pm 0.8\pm 0.6$&$ 4.6\pm 0.5\pm 0.2$&12$\pm 1$& 1.5$\pm$0.2$\pm$0.1\\
$^{16}O$& &$ 6.9\pm 0.6\pm 0.3$&$ 5.8\pm 0.5\pm 0.2$&36$\pm 3$& 1.2$\pm$0.1$\pm$0.1\\ 
\hline\hline
\end{tabular}
\end{center}
\end{table}
The $R_{A}$ values are listed with their statistical and systematic 
uncertainties those listed in the $\pi^+\Sigma^-$ column account for the 
$\Sigma^-$ loss ($\Sigma^-_{loss}$) at momenta around 0 MeV/c. For a given 
nucleus, $\Sigma^-_{loss}$ is calculated as follows: by using the momentum 
distribution of positive sigmas, a loss rate is initially defined by the 
ratio of the number of sigmas in the 0 MeV/c peak to the number of sigmas 
in the bump. The number of sigmas in the 0 MeV/c peak as well as in the 
bump is determined by a global best-fit analysis, where the bump is assumed 
to have a Gaussian line-shape. Then, the loss rate of positive sigmas is 
assigned to negative sigmas. As a final note, the $R_A$ data-set is not 
corrected for the pion attenuation in $A'$ nor for the $\Sigma$-$\Lambda$ 
conversion; therefore, these observables represent the {\em emission rates} 
of the reaction $A(K^-_{stop},\pi^\pm\Sigma^\mp)A'$. In column five the 
$R_{+-}$ ratio (the ratio between $R_{A}(\pi^-\Sigma^+)$ and 
$R_{A}(\pi^+\Sigma^-)$) is reported.

Table 1 also gives an account of the rates published by earlier bubble 
chamber measurements, the  emission rates of $^4He$ \cite{expt:katz} and 
the capture rates of $^{12}C$ \cite{expt:vander-velde}. For the purpose 
of comparison, the $^{12}C$ capture rates are reduced to emission rates 
by means of the numbers discussed in Ref. \cite{expt:vander-velde}. In this 
article, $R_{^{12}C}$=12.7 (15.9) for the $\pi^+\Sigma^-$ ($\pi^-\Sigma^+$) 
channel, which becomes 7.9 (9.9) when the $\Sigma$-$\Lambda$ conversion is 
neglected and finally 7.3 (9.2) when the pion-flux attenuation is not taken 
into account. The ratio between the emission and capture rates $\pi^-\Sigma
^+$/$\pi^+\Sigma^-$ remains the same 1.25$\pm$0.09. The same data on carbon 
are reanalyzed in a second article \cite{expt:vander-velde1}, which 
calculates the capture rates for some $\pi\Sigma$ channels. In this article, 
a capture rate $0.131\pm$0.004 for $\pi^+\Sigma^-$ was found (about the 
same as the in the first article) and $0.294\pm$0.010 for $\pi^-\Sigma^+$ 
(about twice the value previously published) resulting in the ratio 2.24$
\pm$0.12 \cite{expt:vander-velde1}. Such a large value of the capture rate 
for $\pi^-\Sigma^+$ is however based on an emission rate of 9.3, a value 
that can be assessed via the number of events reported in Table I of Ref. 
\cite{expt:vander-velde1}. Such a value is in close agreement with 9.2 
the value given in Table 1.
\begin{center}
{\bf IV. DISCUSSION AND  CONCLUSIONS}   
\end{center}
The analysis aimed at examining the elementary process $K^-_{stop}~p
\rightarrow\pi^\pm\Sigma^\mp$ with $p$ belonging to $A$. The 
experimental method of searching for $\pi\Sigma$ pairs relies on the 
capability of FINUDA to detect all the particles involved in the $K$ 
absorption process, i.e., $K^-_{stop}$ (and the associated $K^+)$, $\pi
^\pm$ and $n$. For $^6Li$, this method assigns a signal-to-background 
ratio to sigmas (detected in coincidence with pions) of $(S/B)_{\Sigma^+
}$=3.6 and $(S/B)_{\Sigma^-}$=4.5. The other targets show similar ratios.

Fig. 3 shows that the $\pi^\pm\Sigma^\mp$ missing mass distributions are 
22-24 MeV/c$^2$ FWHM. These widths are only in part explained by the 
measurement uncertainty. When accounting for the instrumental uncertainty, 
FWHM=7.3 MeV/c$^2$, the missing mass strength is mostly found in the range 
$\sim$10$\pm$11 MeV/c$^2$, where $\sim$10 MeV/c$^2$ is the {\it M} 
peak position above the $M_{[^4He+n]}$ threshold and $\pm$11 MeV/c$^2$ 
FWHM is the intrinsic width of the distribution. This missing energy must 
be compared with the available energy, the kaon rest mass, to realize that 
the $\pi^\pm\Sigma^\mp$ final pairs take away from 96\% to 100\% of the 
energy available to the reaction. The fraction of energy lost is below 
4\%, which can be explained by the energy subtracted by de-excitation 
$\gamma$'s or by unobserved fragments of $A'$ as well as by the energy lost 
by sigmas when traveling through a target. This trait establishes the 
quasi-free nature of the reaction. Such a valuation also indicates that the 
$\pi\Sigma$ kinematics is only slightly altered by final-state interactions. 
The above discussion is related to $^6Li$ but similar conclusions can also 
be drawn for the other nuclei, $^7Li$, $^{9}Be$, $^{13}C$ and $^{16}O$. In 
fact, the missing mass distributions (figures not shown) broaden out to a 
maximum of 20\% when going from $A$=6 to $A$=16.

The $\pi^\pm\Sigma^\mp$ data discussed in this article are the result of 
an exclusive measurement, more than 96\% of the available energy is in 
fact taken by $\pi\Sigma$ pairs. Previous results were obtained from either 
studies of inclusive measurements on $^4He$ \cite{theor:harada} and $^{12}C$ 
\cite{theor:gugelot} or semi-inclusive measurements on a variety of nuclei 
\cite{theor:outa}. For this other class of measurements, there are several 
processes leading to the same final state; in Ref. \cite{theor:outa}, the 
$\pi^-$ momentum spectrum induced by the $(K^-_{stop},\pi^-)$ reaction is 
decomposed into six different processes, which makes the separation of the 
($K^-_{stop},\pi^-\Sigma^+$) process from the others difficult. 

Fig. 4 shows the momentum spectra of prompt pions, which exhibit a 
continuum behavior from the kinematic threshold down to 100 MeV/c. These 
spectra are not corrected for the pion attenuation. Above the kinematic 
threshold (indicated by arrows in Fig. 4), the $\pi^\pm$ momentum spectra 
are seen to be nearly  unaffected by background. The pion spectra are not 
compared with theoretical predictions, although the absorption of 
at-rest kaons in nuclei is addressed in Ref. \cite{theor:dover}. The 
authors develop a plain model of the interaction, which is only capable of 
explaining the shape of the spectra without helping the interpretation of
the data presently discussed. The similarity among the measured momentum 
distributions of sigmas shows that the $K^-_{stop}A$ absorption reaction 
depends weakly on $A$ for p-shell nuclei. This is also corroborated by the 
analyzes based on Monte-Carlo simulations, which show that the quasi-free 
reaction $K^-_{stop}A\rightarrow\pi\Sigma A'$ is capable of explaining the 
shapes of all the distributions. Moreover, the sigma momentum spectra are 
weakly distorted by $\Sigma$-$A$ final-state interactions since they mainly 
proceed via the $\Sigma N\rightarrow\Lambda N$ conversion reaction, which 
selectively removes sigmas from the outgoing flux. Therefore, the M-C 
generated sigma momentum distributions (full dots in Fig. 5) give a fair 
account of the shapes of sigma momenta. The $\Sigma$ to $\Lambda$ conversion 
rate was not determined for these spectra.    

The emission rates  $\pi^\pm\Sigma^\mp$/$K^-_{stop}$ are reported in Table 
1. They are also compared with earlier bubble chamber measurements for 
$^4He$ \cite{expt:katz} and $^{12}C$ \cite{expt:vander-velde}, which were 
published in the early seventies. To this day, these are the only available 
data. When the $K^-_{stop}~p\rightarrow\pi^\pm\Sigma^\mp$ elementary process 
occurs on a bound proton the relationship $R_{+-} >$1 holds for light 
nuclei; in fact, the $R_{+-}$ ratio ranges from 1.8 to 1.3 to 1.2 for $^4
He$, $^9Be$ and $^{16}O$, respectively. If the same process occurs on a 
free proton ($H$) then the relationship is reversed: $R_{+-}$=0.42 ($\gamma 
^{-1}$ in Ref. \cite{expt:novak}). Such a clear increase of $R_{+-}$, when 
moving the interaction from a free proton to a bound proton, can be 
related to the sub-threshold modification of the $\bar{K}N$ interaction. 

For the light nuclei examined, the average value of $R_{+-}$ is 1.3$\pm$0.1 
whose flat behavior indicates that the kaon absorption occurs preferentially
at the surface of a nucleus. This value of $R_{+-}$ can be compared directly
with the ratio $R_{+-}$=Yield $\pi^-\Sigma^+$/Yield $\pi^+\Sigma^-$ being 
calculated in Ref. \cite{theor:wycech}. In this article, the behavior of 
$R_{+-}$ is given as a function of $\rho/\rho_0$, where $\rho_0$ is the 
nuclear density at saturation. $R_{+-}$ is predicted to vary from 1.4 to 
1.7 for $\rho/\rho_0$ ranging from 0.0 to 0.35, this range accounting for 
the nuclear skin densities of light nuclei. Moreover, the ratio $R_{+-}$ 
corresponds to the model {\em c} option, that is, the option that accounts 
for final state particle absorption. The agreement between theoretical 
predictions and the experimental value of $R_{+-}$ is remarkable. The 
discussion on the measured emission rates cannot take further advantage of 
the comparison with theory because of the lack of model calculations. The 
exception is the inclusive reaction $^{12}C(K^-_{stop}~,\pi^\pm)$, where 
capture rates as well as pion momentum spectra are calculated 
\cite{theor:ohnishi}. 

The absorption of stopped kaons involves only surface protons when the
final state particles are $\pi^\pm\Sigma^\mp$ pairs. Nevertheless, Table 
1 shows that when a neutron is added to a nucleus $R_A$ drops for both 
channels; in fact, for the lithium isotopes is $R_{^7Li}/R_{^6Li}$=0.44 
for both $\pi^-\Sigma^+$ and  $\pi^+\Sigma^-$. This fact reflects the 
excess of surface neutrons over surface protons of $^7Li$ with respect to 
$^6Li$. Such an excess has the net effect of depleting the number of $K^-A$ 
stop centers leading to $\pi^\pm\Sigma^\mp$ pairs. The same relationship is
 also found for the carbon isotopes, $R_{^{13}C}/R_{^{12}C}$=0.75 for $\pi
^-\Sigma^+$ and 0.63 for $\pi^+\Sigma^-$. As a final note, the large values 
of $R_A$'s found for $^6Li$ reflect the previously measured large value for 
$^4He$ since this nucleus forms the lithium core. 
\begin{center}
{\bf ACKNOWLEDGMENTS}
\end{center}
This work was supported by the Istituto Nazionale di Fisica Nucleare (INFN) 
of Italy. It is a pleasure to thank Prof. A. Gal for the useful suggestions. 
The support received from the DA$\Phi$NE crew was indispensable for the 
success of the experiment, which was highly appreciated.
%
%

%
%

\begin{thebibliography}{99}
         %
         %
\bibitem{expt:agnello1}
M. Agnello et al., Phys. Lett. {\bf B622} (2005) 53.
%
\bibitem{expt:agnello2}
M. Agnello et al., Phys. Lett. {\bf B698} (2011) 219.
%
\bibitem{theor:cieply}
A. Cieply et al., Phys. Lett. {\bf B698} (2011) 226.
%
\bibitem{expt:agnello3}
M. Agnello et al., Phys. Lett. {\bf B681} (2009) 139.
%
\bibitem{expt:agnello4}
M. Agnello et al., Nucl. Phys. {\bf A804} (2008) 151.
%
\bibitem{expt:agnello5}
M. Agnello et al., Phys. Lett. {\bf B685} (2010) 247.
%
\bibitem{expt:agnello6}
M. Agnello et al., Phys. Rev. Lett. {\bf 94} (2005) 212303.
%
\bibitem{expt:agnello7}
M. Agnello et al., Phys. Lett. {\bf B654} (2007) 80.
%
\bibitem{expt:agnello8}
M. Agnello et al., Phys. Lett. {\bf B669} (2008) 229.
%
%
\bibitem{theor:weise-gal}
W. Weise, Nucl. Phys. {\bf A835} (2010) 51.             \\
A. Gal, Prog. Theor. Phys. Suppl. {\bf 186} (2010) 270.
%
\bibitem{theor:wycech}
S. Wycech, Nucl. Phys. {\bf B28} (1971) 541.
%
\bibitem{theor:staronski}
L. R. Staronski and S. Wycech, J. Phys. {\bf G13} (1987) 1361
%
\bibitem{theor:ohnishi}
A. Ohnishi, Y. Nara and V. Koch, Phys. Rev. {\bf C56} (1997) 2767.
%
\bibitem{theor:friedman}
E. Friedman and A. Gal, Phys. Rep.  {\bf 452} (2007) 89.
%
\bibitem{theor:oset}
D. Jido, E. Oset and T. Sekihara, Eur. Phys. Jour. {\bf A47} (2011) 42.
%
         %
         %
\bibitem{FINUDA:00}
V. Filippini, M. Marchesotti, and C. Marciano,
{\it Nucl. Instr. and Methods} {\bf A424} (1999), 343.
%
\bibitem{FINUDA:01}
A. Pantaleo et al., Nucl. Instr. and Methods {\bf A545} (2005) 593.
%
\bibitem{expt:agnello9} 
M. Agnello et al., Phys. Lett. {\bf B701} (2011) 556.
%
\bibitem{FINUDA:02}
P. Bottan et al., Nucl. Instr. and Methods {\bf A427} (1999) 423.
%
\bibitem{FINUDA:03}
M. Agnello et al., Nucl. Instr. and Methods {\bf A385} (1997) 58.
%
\bibitem{FINUDA:04}
L. Benussi et al., Nucl. Instr. and Methods {\bf A361} (1995) 180; \\
%
L. Benussi et al., Nucl. Instr. and Methods {\bf A419} (1998) 648.
%
%
\bibitem{expt:katz}
P.A. Katz et al., Phys. Rev. {\bf D1} (1970) 1267.
%
\bibitem{expt:vander-velde}
C. Vander Velde-Wilquet et al., Nucl. Phys. {\bf A241} (1975) 511.
%
\bibitem{expt:vander-velde1}
C. Vander Velde-Wilquet et al., Nuovo Cim. {\bf 39 A} (1977) 538.
%
\bibitem{theor:harada}
T. Harada and T. Akaishi Phys. Lett. {\bf B262} (1991) 205.
%
\bibitem{theor:gugelot}
P. C. Gugelot, S. M. Paul and R. D. Ransome, Phys. Rew. {\bf C41} (1990) 2445.
%
\bibitem{theor:outa}
H. Outa et al., Progr. of Theor. Phys. Suppl. {\bf 117} (1994)  177.
%
\bibitem{theor:dover}
C. B. Dover, D. J. Millener and A. Gal, Phys. Rep.  {\bf 184} (1989) 1.
%
\bibitem{expt:novak}
R. J. Nowak et al., Nucl. Phys. {\bf B139} (1978) 61.
%
\end{thebibliography}
\end{document}